\documentclass[twocolumn,a4paper]{article}  
\usepackage[twocolumn,textwidth=18cm,columnsep=.81cm]{geometry}
\RequirePackage{graphicx}
\usepackage{amssymb}
\usepackage{amsmath}
\usepackage{epstopdf}
\usepackage{array}
\newcolumntype{L}[1]{>{\raggedright\let\newline\\\arraybackslash\hspace{0pt}}m{#1}}
\newcolumntype{C}[1]{>{\centering\let\newline\\\arraybackslash\hspace{0pt}}m{#1}}
\newcolumntype{R}[1]{>{\raggedleft\let\newline\\\arraybackslash\hspace{0pt}}m{#1}}
\usepackage[table]{xcolor}
\usepackage{booktabs}
\usepackage{multirow}
\usepackage[colorlinks,citecolor=red,urlcolor=blue,bookmarks=false,hypertexnames=true]{hyperref} 
\usepackage{float}
\usepackage[]{subfig}
\colorlet{tableheadcolor}{gray!25}
\colorlet{tablerowcolor}{gray!12.5}
\usepackage{tablefootnote}
\usepackage{siunitx}
\usepackage{lineno}
\DeclareMathOperator\erf{erf}
\DeclareMathOperator\erfc{erfc}

\title{Estimating the efficiency turn-on curve for a constant-threshold trigger without a calibration dataset.}

\begin{document}
\author{T.R. Pollmann$^1$}


\date{%
    $^1$Department of Physics, Technische Universit\"at M\"unchen, 80333 Munich, Germany\\%
    tina.pollmann@tum.de\\
}

\maketitle
\begin{abstract}
Many particle physics experiments use constant threshold triggers, where the trigger threshold is in an online estimator that can be calculated quickly by the trigger module. Offline data analysis then calculates a more precise offline estimator for the same quantity, for example the event energy. The efficiency curve is a step function in the online estimator, but not in the offline estimator. 

One typically obtains the shape of the efficiency curve in the offline estimator by way of a calibration dataset, where the true rate of events at each value of the offline estimator is measured once and compared to the rate observed in the physics dataset. For triggers with a fixed threshold condition, it is sometimes possible to bootstrap the trigger efficiency curve without use of a calibration dataset. This is useful to verify stability of a calibration over time when calibration data cannot be taken often enough. It also makes it possible to use datasets for which no calibration is available. This paper describes the method and the conditions that must be met for it to be applicable.
\end{abstract}

\section{Introduction}

In many particle physics experiments, the data acquisition system (DAQ) monitors the signals from the detector, waiting for an event of interest to occur. Once such an event does happen, the DAQ has only a very short time window in which to notice the event and trigger data readout. Hence, the decision about when to trigger must be based on only a rough estimate of the parameter of interest (such as the event energy), which can be calculated with the required speed\cite{Harvey:1990in}.

Consider a situation where the DAQ bases the trigger decision on an online estimator $x$. Data read-out is triggered whenever $x \geqslant \theta_x$, so the trigger efficiency in $x$ is a step function $\epsilon(x) = \Theta(x-\theta_x)$. In offline data analysis, the estimator $y$ is calculated. $y$ is a better estimator for the variable under consideration (e.g. the event energy or momentum) as it is determined using more elaborate algorithms and calibration inputs. Of interest for analysis is the trigger efficiency as a function of $y$, $\epsilon(y)$.

A number of methods exist to determine $\epsilon(y)$ by way of a calibration dataset. These rely on measuring directly or indirectly the true rate of events at each value of $y$, so that by comparison with the rate obtained after the trigger, the trigger efficiency can be calculated.

Obtaining a trigger efficiency calibration is not always possible. The calibration data could be corrupt, it could be impossible to record calibration data due to electronics or physics constraints, or the calibration could drift with time faster than calibration datasets can be taken. In such cases, the efficiency curve might still be recovered or verified provided the value of $x$ for each event was recorded or can be obtained offline.
 
\section{General principle and illustration of the method}
A dataset contains the value of $x$ and $y$ for each event. For concreteness, we say that these are both estimators for the event energy. We assume that the events recorded have a continuous spectrum in both $x$ and $y$ in the region relevant to the trigger.

Consider the histogram of $x$ versus $y$ for each event, $I(x,y)$. Because $x$ and $y$ are estimators for the same quantity, they are correlated and the data will form a `band' in this 2-dimensional space. The data has a spectrum in the $x$ parameter, $I(x) = \int_{-\infty}^\infty I(x,y) dy$, and a spectrum in the y parameter $I(y) = \int_{-\infty}^\infty I(x,y) dx$

\begin{figure}[htb]
\begin{center}
\subfloat[]{\includegraphics[width=0.75\columnwidth]{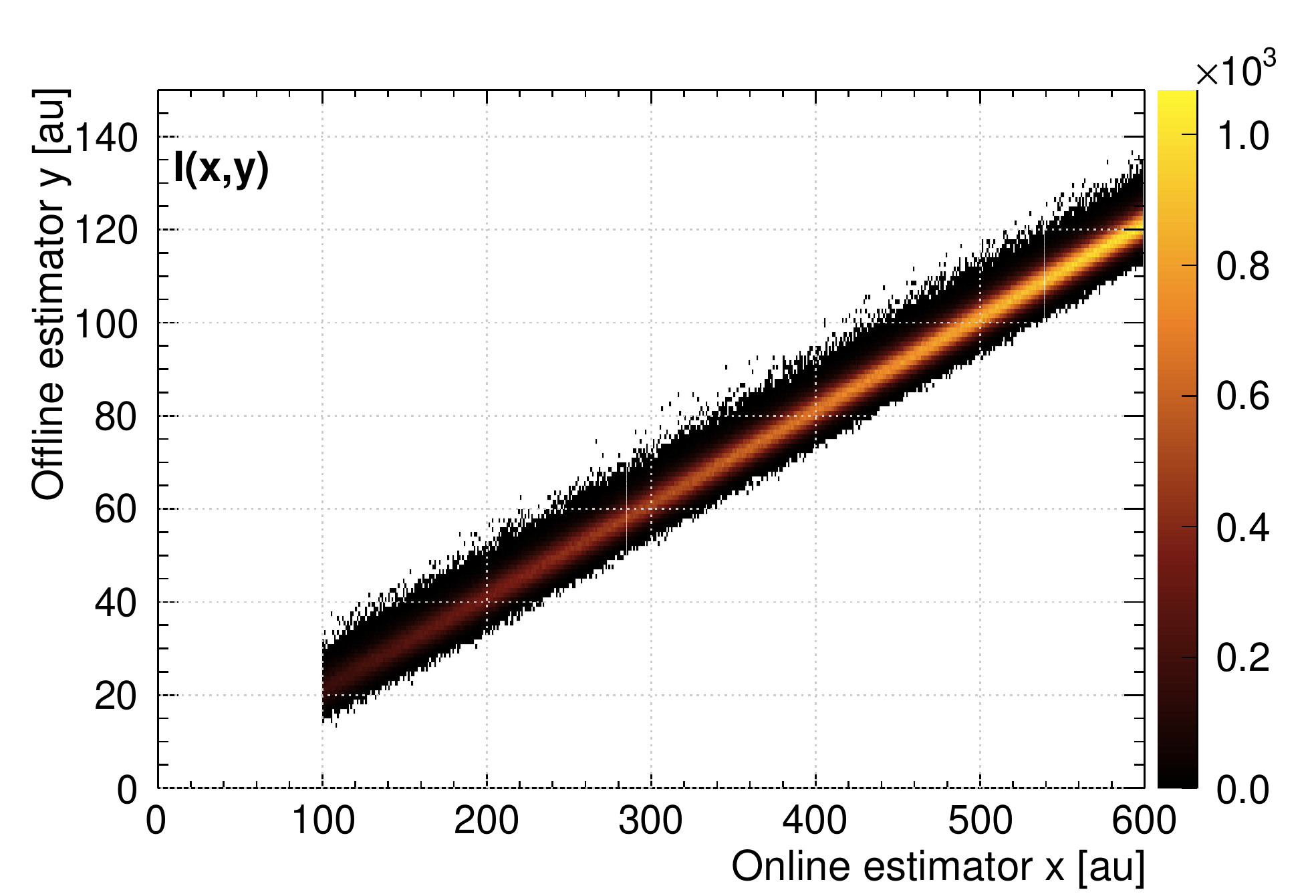}}\\
\subfloat[]{\includegraphics[width=0.75\columnwidth]{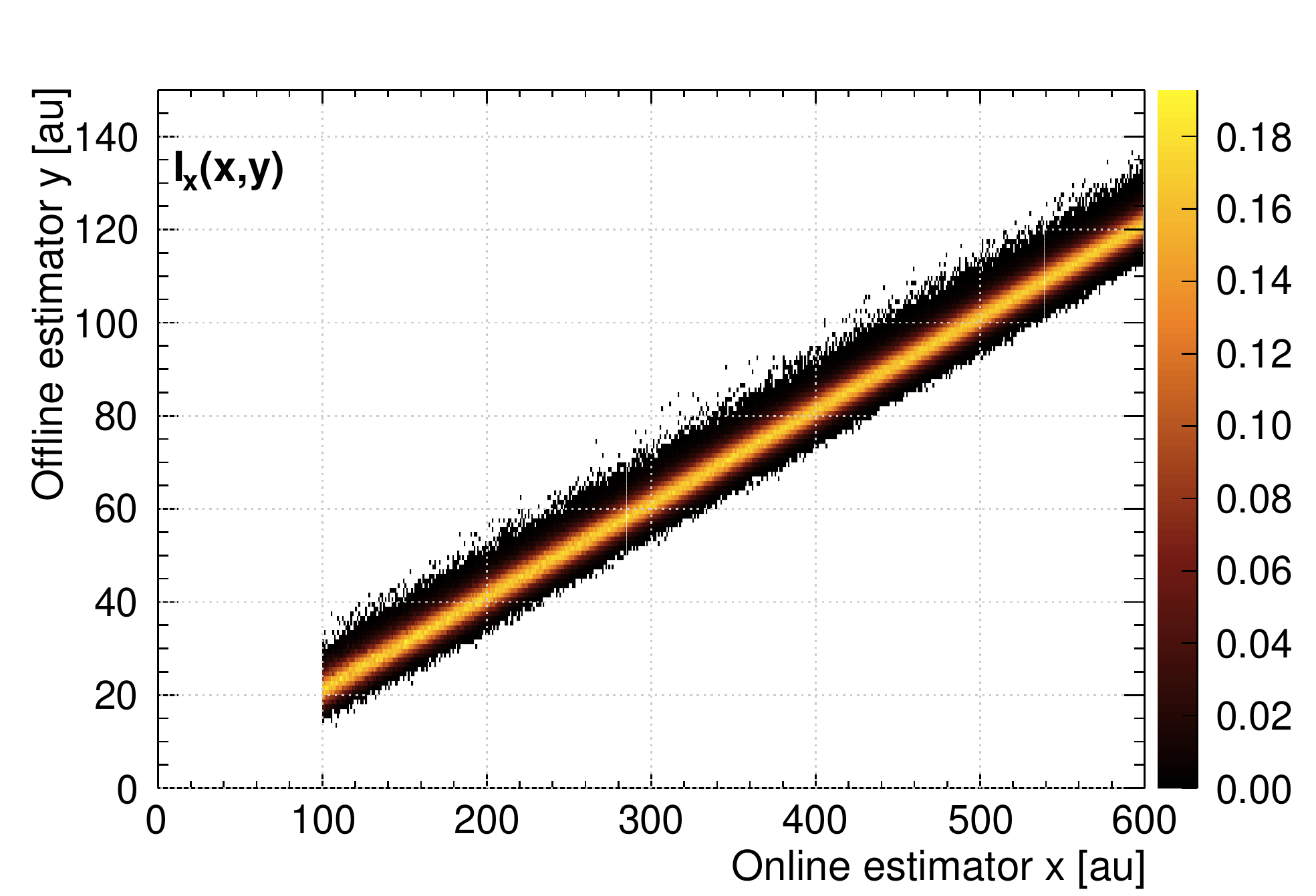}}
\caption{Simulated data using a skewed Gaussian resolution function with $\mu = 0.2x$, $\sigma = 0.2$, and $\lambda = 0.7$ (see Eq.~\eqref{eq:skewedg}). a) The $I(x,y)$ distribution as it might be measured in an experiment. b) The $I_{x}(x,y)$ distribution, where each bin in $I(x,y)$ is normalized such that $I(x) = 1$ for $x \geqslant \theta_x$.}
\label{fig:skewgausmc1}
\end{center}
\end{figure}

To illustrate how to obtain $\epsilon(y)$ from a dataset, we create data in a toy Monte Carlo (MC) simulation with a spectrum $I(x) = 10x$, a trigger threshold $\theta_x = 100$, and a resolution such that the shape of the $y$ distribution for events of the same $x$ is a skewed Gaussian. The $I(x,y)$ histogram for the simulated events is shown in Fig.~\ref{fig:skewgausmc1}~a). The functional form of the relation between $x$ and $y$ is not typically known a-priori in real data, and the method developed here does not rely on such knowledge, so we limit ourselves in this section to information that can be obtained from the data. This situation will be analysed mathematically in Sect.~\ref{sec:skewgauss}.

\begin{figure}[htb]
\begin{center}
\subfloat[]{\includegraphics[width=0.75\columnwidth]{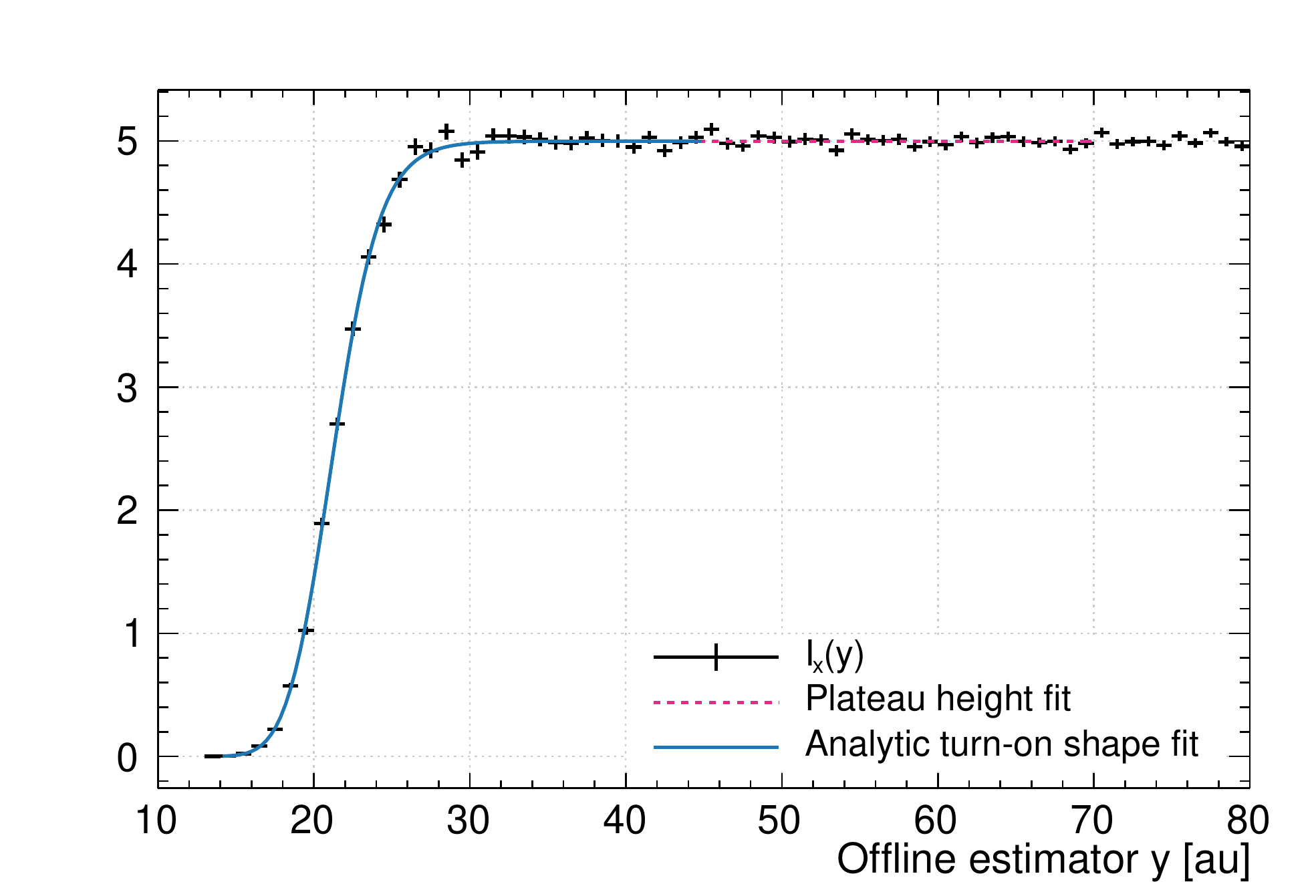}}\\
\subfloat[]{\includegraphics[width=0.75\columnwidth]{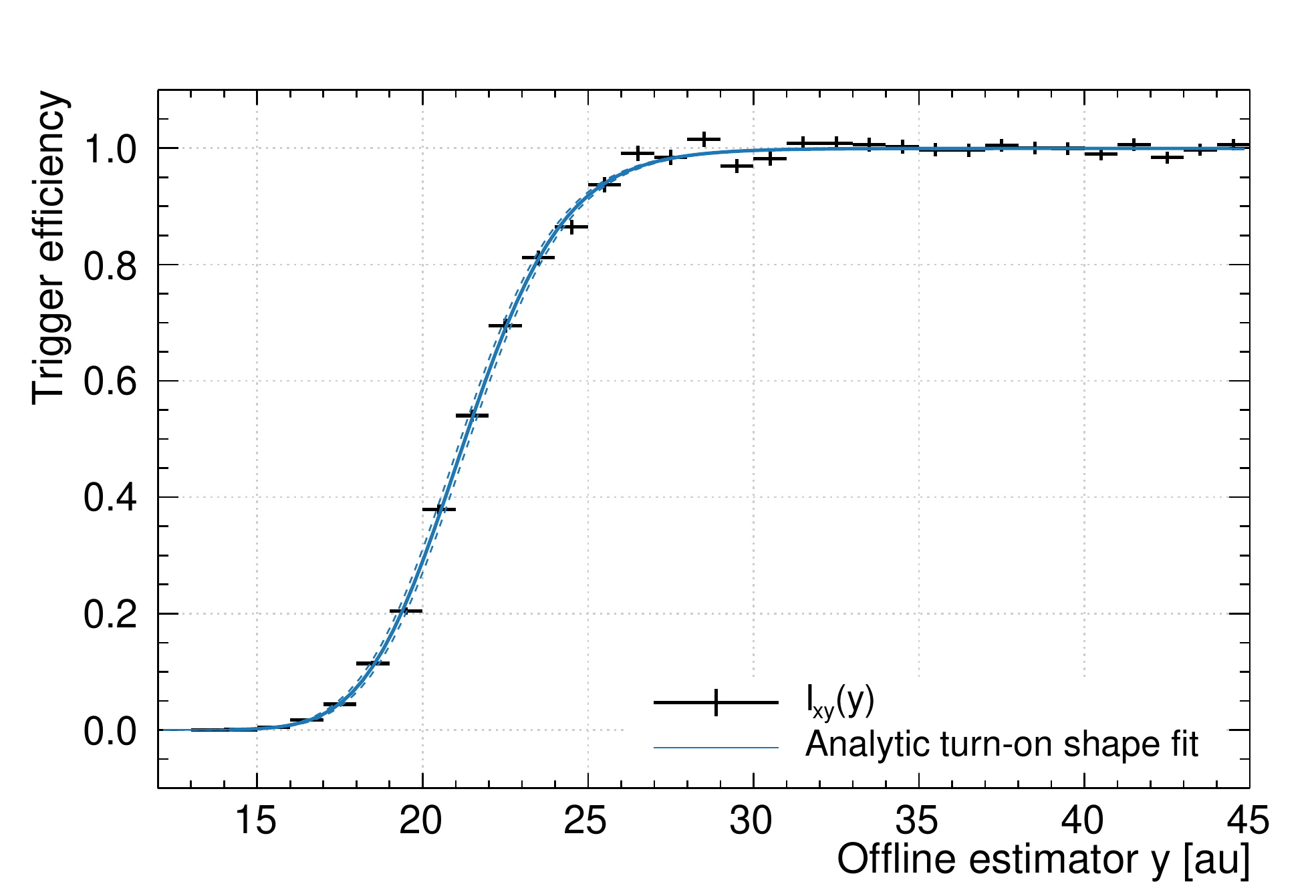}}
\caption{For the same MC data as Fig.~\ref{fig:skewgausmc1} a) The $I_x(y)$ distribution is fit with a straight line in a region far away from the trigger threshold, and with the analytic description of the turn-on shape, which is known in this case. d) $I_x(y)$ scaled by the plateau height results in the trigger efficiency curve (zoomed in compared to panel a) to better show the turn-on region).}
\label{fig:skewgausmc2}
\end{center}
\end{figure}

Fig.~\ref{fig:skewgausmc1}~a) shows what the real detector data might look like. To obtain the trigger turn-on curve in $y$, the following steps are taken

\begin{enumerate}
\item Normalize the $I(x,y)$ histogram such that $I(x) = 1$ for $x \geqslant \theta_x$. This can be achieved by dividing each bin (x,y) in the histogram by $I(x)$. We denote the histogram normalized in $x$ as $I_x(x,y)$. It is shown in Fig.~\ref{fig:skewgausmc1}~b). After this normalization, $I_x(x)$ is equal to the efficiency curve $\epsilon(x) = \Theta(x-\theta)$. 
\item Now consider the spectrum in $y$, that is $I_x(y) = \int_{-\infty}^\infty I_x(x,y) dx$, illustrated in Fig.~\ref{fig:skewgausmc2}~a). For values of $y$ where all possible values of $x$ are above the trigger threshold (approximately at $y\geqslant30$ in this example), a constant plateau arises. For values of $y$ where some of the possible $x$ values are below the trigger threshold, the spectrum is diminished by that fraction of $x$ values which lies below the threshold. 
\item The height of the plateau is determined by a fit with a straight line (red dashed line in Fig.~\ref{fig:skewgausmc2}~a)).
\item Finally, $I_x(y)$ is divided by the plateau height. The resulting curve, shown in Fig.~\ref{fig:skewgausmc2}~b), is an estimate of the trigger efficiency turn-on function.
\end{enumerate}

Because in the simulation the functional dependence is known, the functional shape of the turn-on curve can be fit to the data. This is the blue solid line in Fig.~\ref{fig:skewgausmc2}~a) and b).

%

The crucial feature necessary for this method to work is the constant plateau in $I_x(y)$. In the most general case, $I_x(y)$ can have a nearly arbitrary shape. Under some conditions though, $I_x(y)$ does not depend on $x$ or $y$ until the trigger threshold is introduced. That is, $\hat{I}_x(y) = c$ where the hat indicates the absence of the trigger and $c$ is a constant plateau height. Any value in the actual $I_x(y)$ histogram not equal to $c$ then indicates the influence of the trigger and the difference $c - I_x(y)$ is proportional to the number of events missing at $y$ due to the trigger efficiency\footnote{This assumes that we know that the trigger efficiency tends toward exactly $1$. A value different from $1$ can be trivially accommodated in the scaling done in the last step.}

\section{Validation}

In the following two sections, we analytically demonstrate the validity of this method for two common resolution functions and discuss the conditions that must be met to obtain a constant plateau region.

\subsection{Gaussian example}

In this section, the method of determining the trigger efficiency is discussed mathematically for Gaussian resolution functions. The offline estimator $y$ follows a Gaussian distribution for any given value of $x$. The shape parameters of the Gaussian distribution are functions of $x$, and the data has some spectrum $N(x)$\footnote{We explicitely name the spectrum $N(x)$ here. By integration, we find that $I(x) = \int I(x,y)dy = N(x)$.}:
\begin{linenomath*}
\begin{align}
I(x,y) = \begin{cases}
N(x) \frac{1}{\sqrt{2\pi}\sigma(x)} e^{-(y-\mu(x))^2/(2\sigma(x)^2)} &(x \geqslant \theta_x)\\
       0 &(x < \theta_x)
       \end{cases}
\end{align}
Division by $N(x)$ gives $I_x(x,y)$ which is by construction already normalized such that $I_x(x) = 1$ above the trigger threshold. To obtain $I_x(y)$, an assumption must be made about the shape parameters. In the simplest case, $\sigma(x) = \sigma$ and $\mu(x) = a\cdot x$. Thus:
\begin{align}
I_x(x,y) = \begin{cases}
		 \frac{1}{\sqrt{2\pi}\sigma} e^{-(y-ax)^2/(2\sigma^2)} &(x \geqslant \theta_x) \\ \label{eq:toygaus2d}
        0  &(x < \theta_x)
       \end{cases}
\end{align}
\end{linenomath*}

We temporarily ignore the trigger condition to study the spectrum in $y$ (indicated by the hat). 
\begin{linenomath*}
\begin{align}
\hat{I}_x(y) &= \int_{-\infty}^\infty \frac{1}{\sqrt{2\pi}\sigma} e^{-(y-a\cdot x)^2/(2\sigma^2)} dx \\
     &= \frac{1}{a}
\end{align}
\end{linenomath*}

We find that the spectrum in $y$ is a constant if no trigger condition is applied. Thus, we proved here that the critical condition for the method to work is met, i.e. that for values of y well above the trigger region, $I_x(y)$ forms a constant plateau. 

The analytic shape of the turn-on curve can be obtained by including the trigger condition in the integral
\begin{linenomath*}
\begin{align}
I_x(y) &= \int_{-\infty}^\infty \frac{1}{\sqrt{2\pi}\sigma} e^{-(y-a\cdot x)^2/(2\sigma^2)} H(x-\theta_x)dx \label{eq:gausstepconvolution}  \\
     &= \frac{1}{2a}[ \erf{\big(\frac{y-a\theta_x}{\sqrt{2}\sigma}\big)} + 1 ]  \label{eq:gausscaleefficiency}
\end{align}
The integral in Eq.~\eqref{eq:gausstepconvolution} is formally equal to a convolution of a Gaussian with a step function. This curve describes the fraction of events that pass the trigger for each value of $y$, relative to some plateau height $\frac{1}{a}$ that is reached for $y \gg a\theta_x$. It can be turned into the trigger turn-on curve by scaling such that the plateau is at 1
\begin{align}
\epsilon(y) &= I_x(y)\cdot a \\
            &= \frac{1}{2}[ \erf{\big(\frac{y-a\theta_x}{\sqrt{2}\sigma}\big)} + 1 ]  \label{eq:gausefficiency}
\end{align}
\end{linenomath*}

\begin{figure*}[h]
\begin{center}
\subfloat[]{\includegraphics[width=0.75\columnwidth]{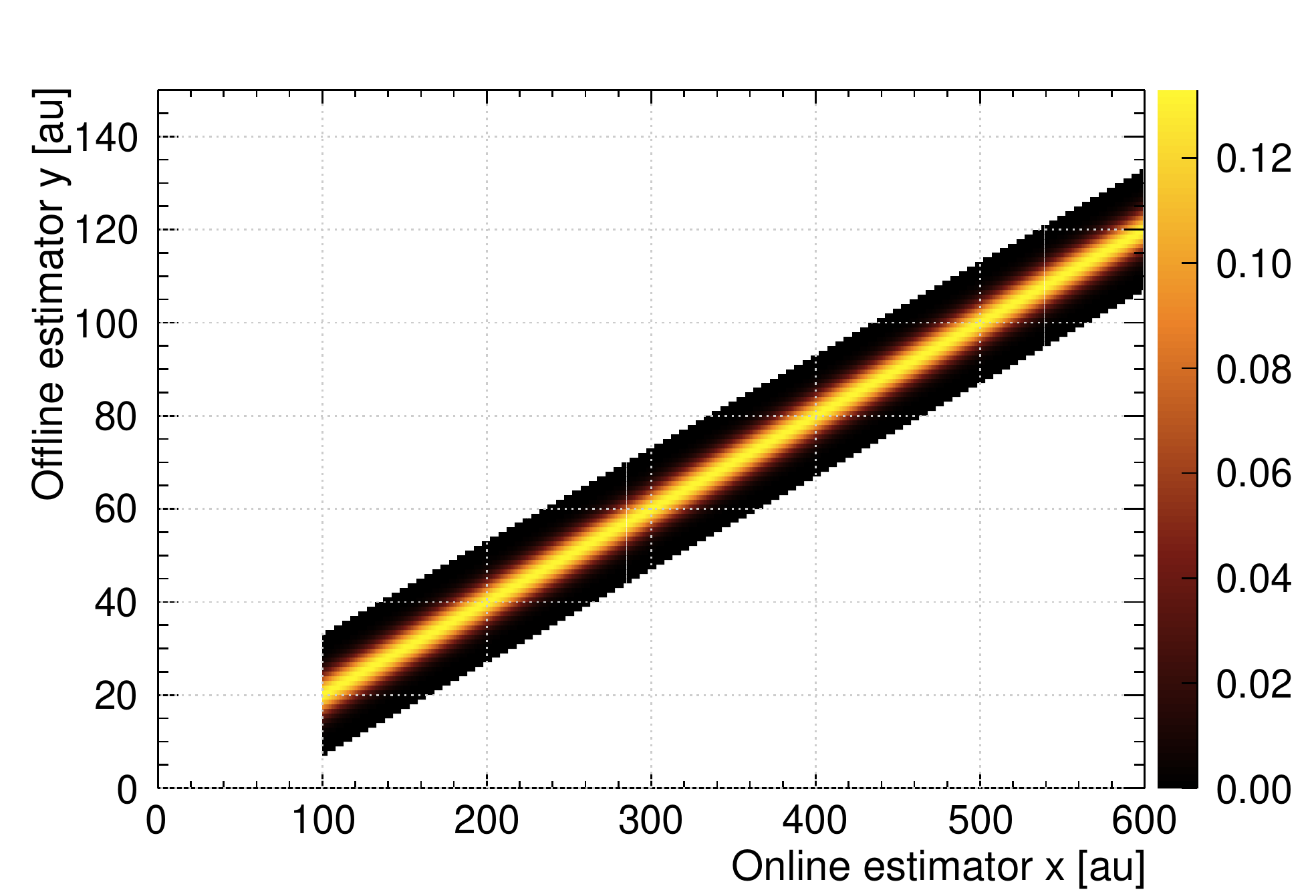}}\\
\subfloat[]{\includegraphics[width=0.75\columnwidth]{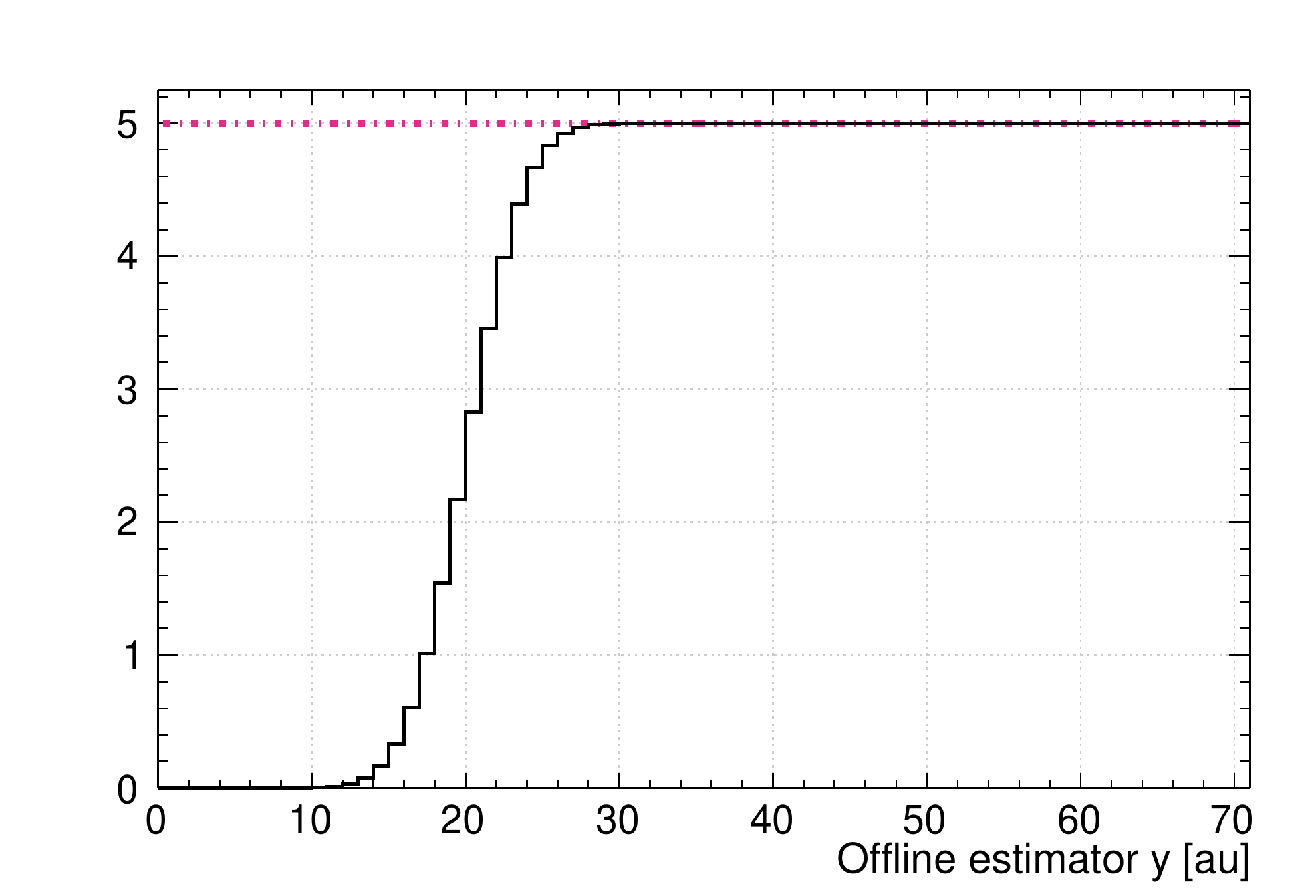}}
\subfloat[]{\includegraphics[width=0.75\columnwidth]{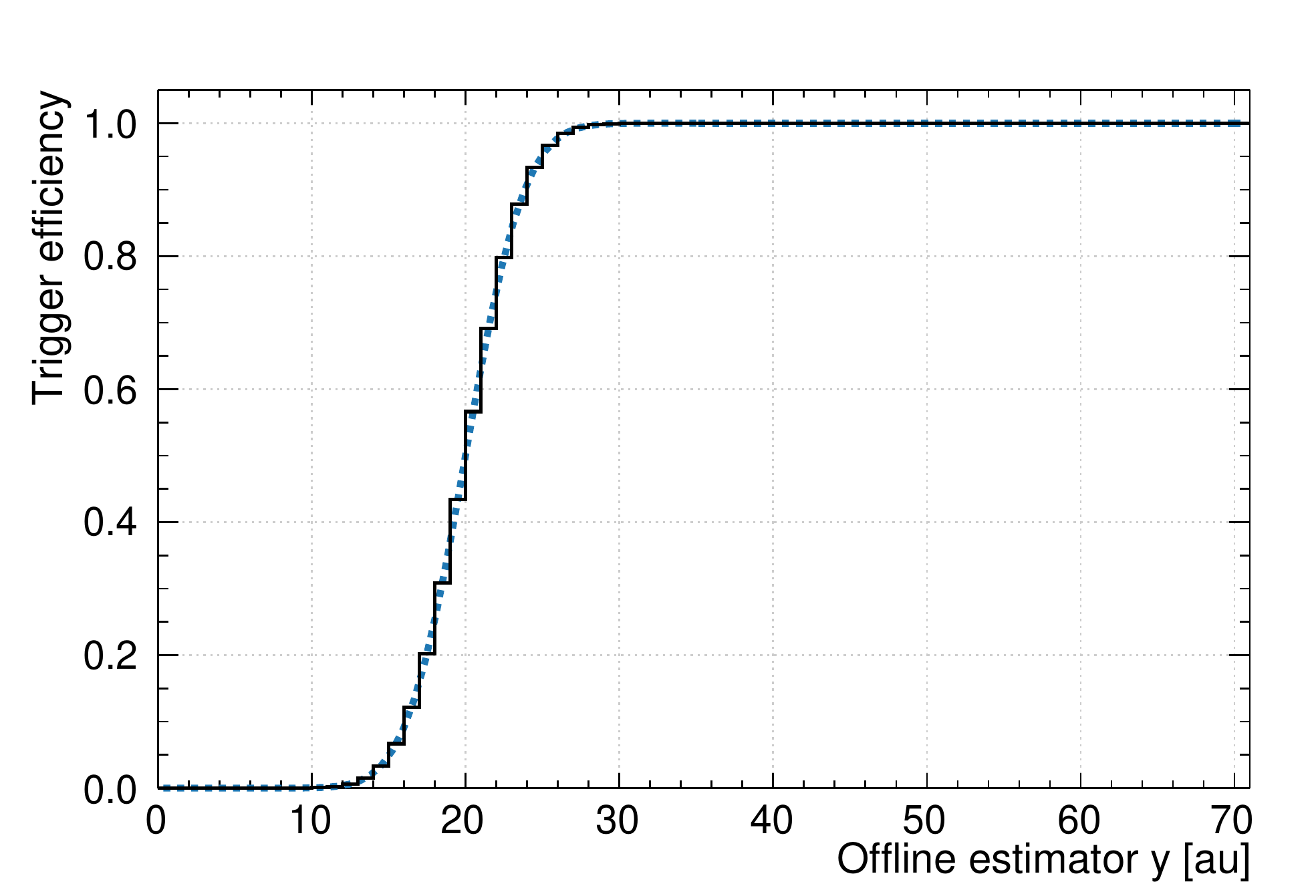}}
\caption{The Gaussian example (Eq.~\eqref{eq:toygaus2d}) for $a=0.2$, $\sigma = 3$. a) The distribution of $x$ versus $y$. b) The y-profile  (black line), together with the calculated level of the plateau (pink dashed). c) The profile scaled by the plateau height (black), which represents the trigger efficiency. The blue dashed line is a step function convoluted by a Gaussian, with function parameters taken at the trigger threshold.}
\label{fig:toygaus1d}
\end{center}
\end{figure*}

Fig.~\ref{fig:toygaus1d} illustrates Eq.~\eqref{eq:toygaus2d} (panel a), Eq.~\eqref{eq:gausscaleefficiency} (panel b), and Eq.~\eqref{eq:gausefficiency} (panel c). Parameters used are $a=0.2$, $\sigma = 3$ and $\theta_x = 100$. This is not a Monte Carlo simulation; the respective equations are evaluated numerically here.

\begin{figure}[htb]
\begin{center}
\subfloat[]{\includegraphics[width=0.75\columnwidth]{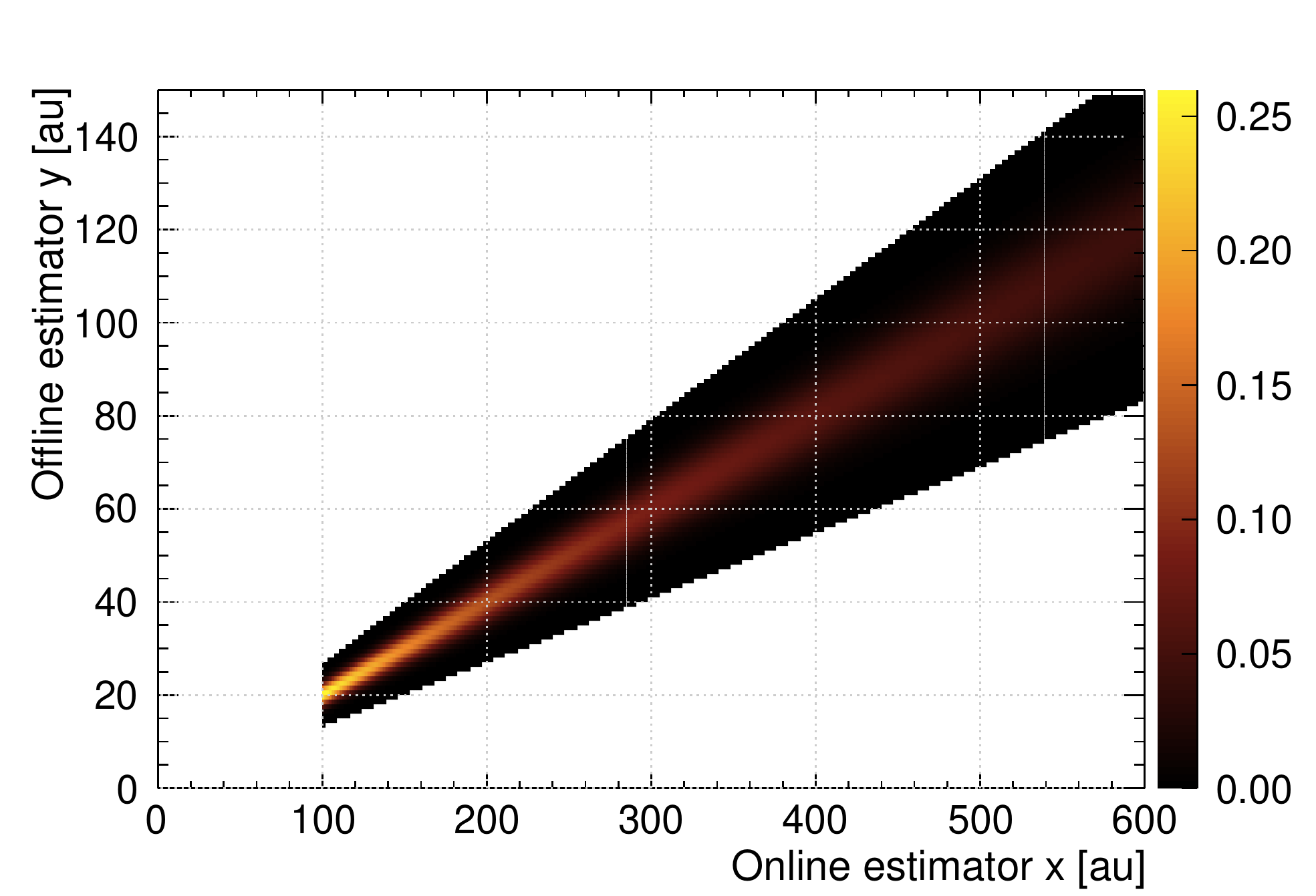}}\\
\subfloat[]{\includegraphics[width=0.75\columnwidth]{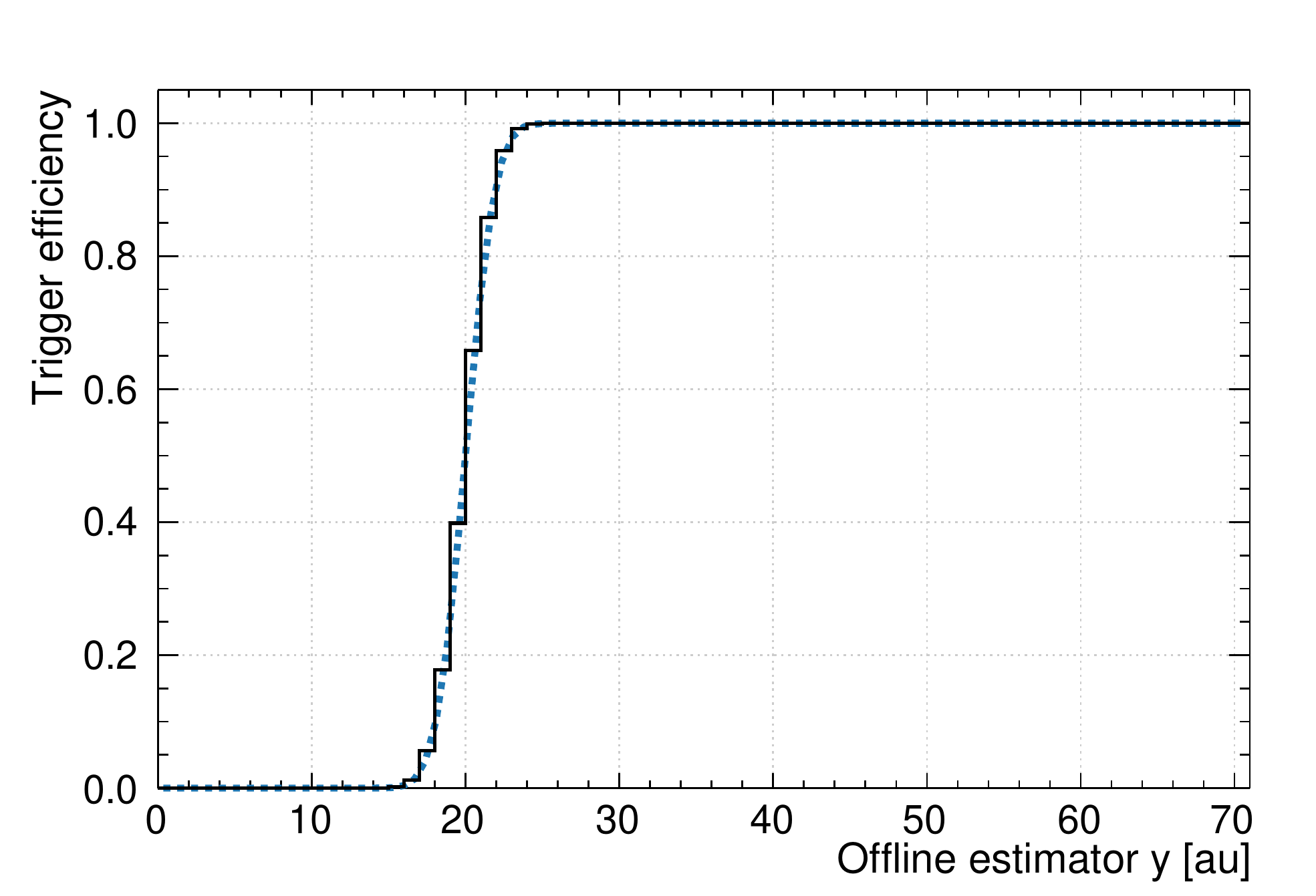}}
\caption{The Gaussian example with $\mu = 0.2y$ and $\sigma = 0.015y$. a) $x$ vs $y$. b) The efficiency curve in $y$.}
\label{fig:gausmodelm2s2}
\end{center}
\end{figure}

In a more realistic case, both the mean and the width of the $y$ distribution at a given $x$ vary with $x$: $\sigma(x) = b\cdot x$ and $\mu(x) = a\cdot x$ so that
\begin{linenomath*}
\begin{align}
I_x(y) & = \int_{-\infty}^\infty \frac{1}{\sqrt{2\pi}bx} e^{-(y-a\cdot x)^2/(2(bx)^2)} dx \\
\end{align}
\end{linenomath*}
This integral cannot be solved analytically. The numeric solution for $a = 0.2y$ and $b = 0.015y$ is shown in Fig.~\ref{fig:gausmodelm2s2}. Panel (b) shows that a constant plateau exists and the plateau height is determined by a fit to the histogram between 30 and 50~y. The trigger turn-on curve in panel (b) is overlaid with the model from Eq.~\eqref{eq:gausefficiency} with $\sigma = \sigma(\theta_x)$.

\begin{figure}[htb]
\begin{center}
\subfloat[]{\includegraphics[width=0.75\columnwidth]{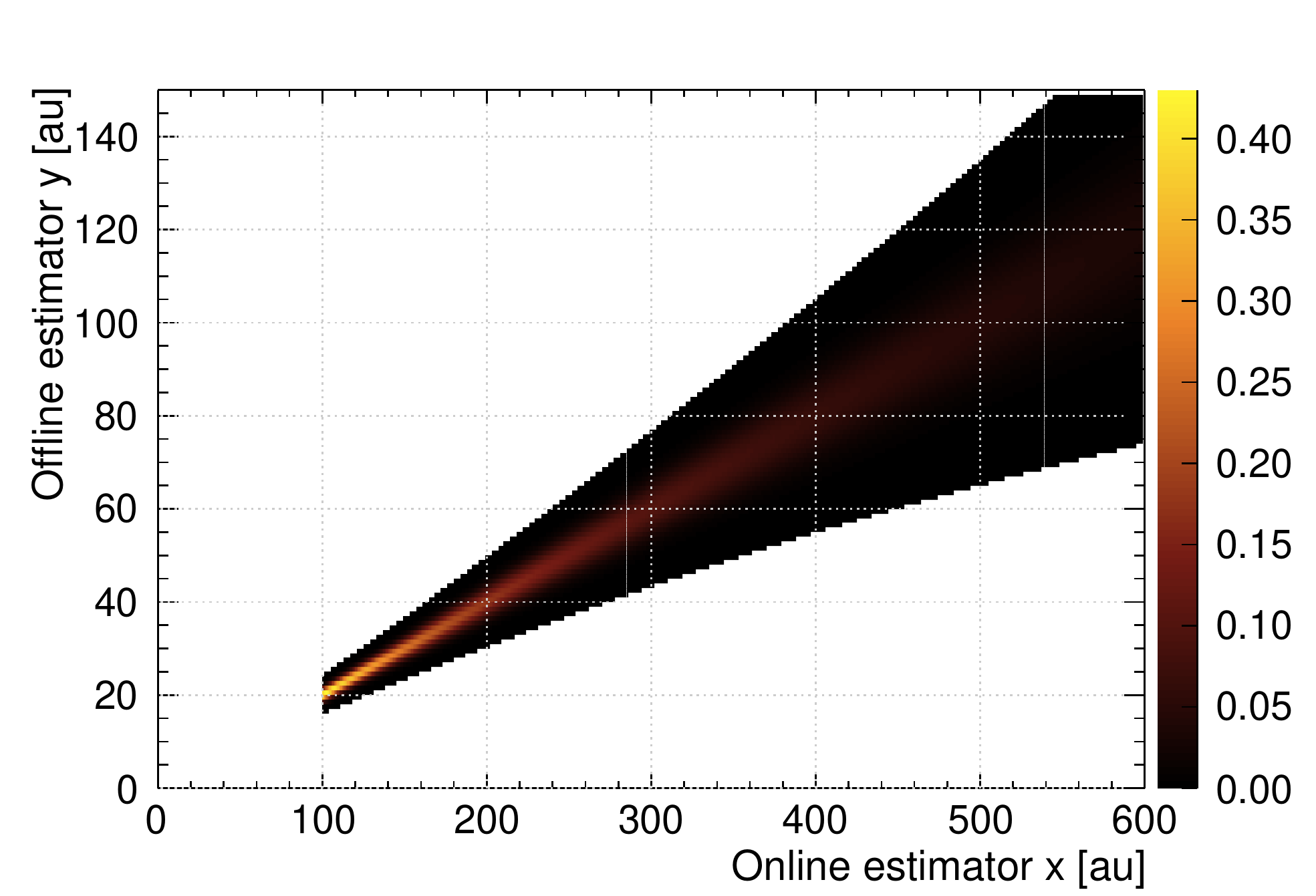}}\\
\subfloat[]{\includegraphics[width=0.75\columnwidth]{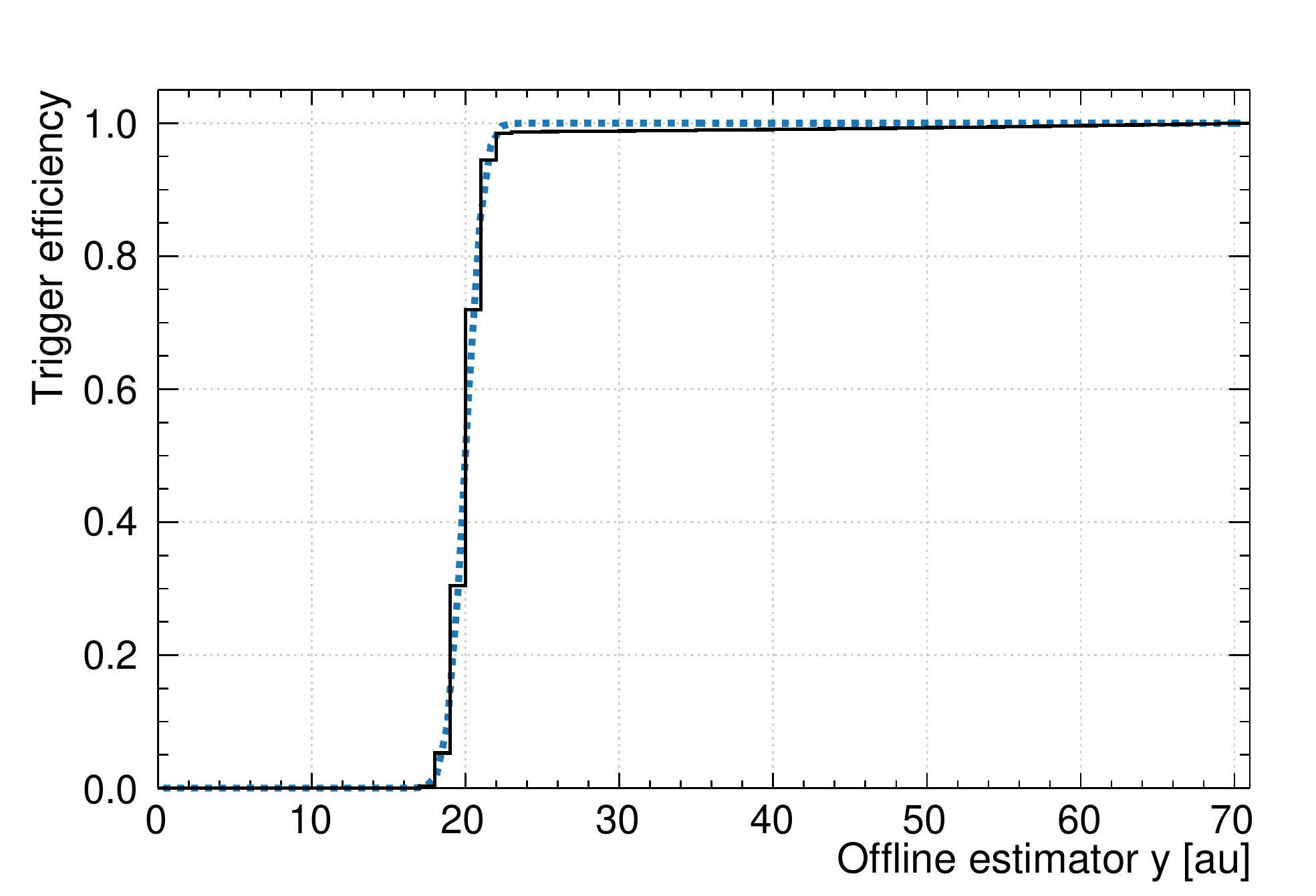}}
\caption{The Gaussian example with $\mu = 0.2y$ and $\sigma = 0.07y + 2\cdot10^{-5} y^2$. No constant plateau arises, hence the method is not applicable and panel b) does not represent the trigger efficiency.}
\label{fig:gausmodelm2s3}
\end{center}
\end{figure}

\begin{figure}[htb]
\begin{center}
\subfloat[]{\includegraphics[width=0.75\columnwidth]{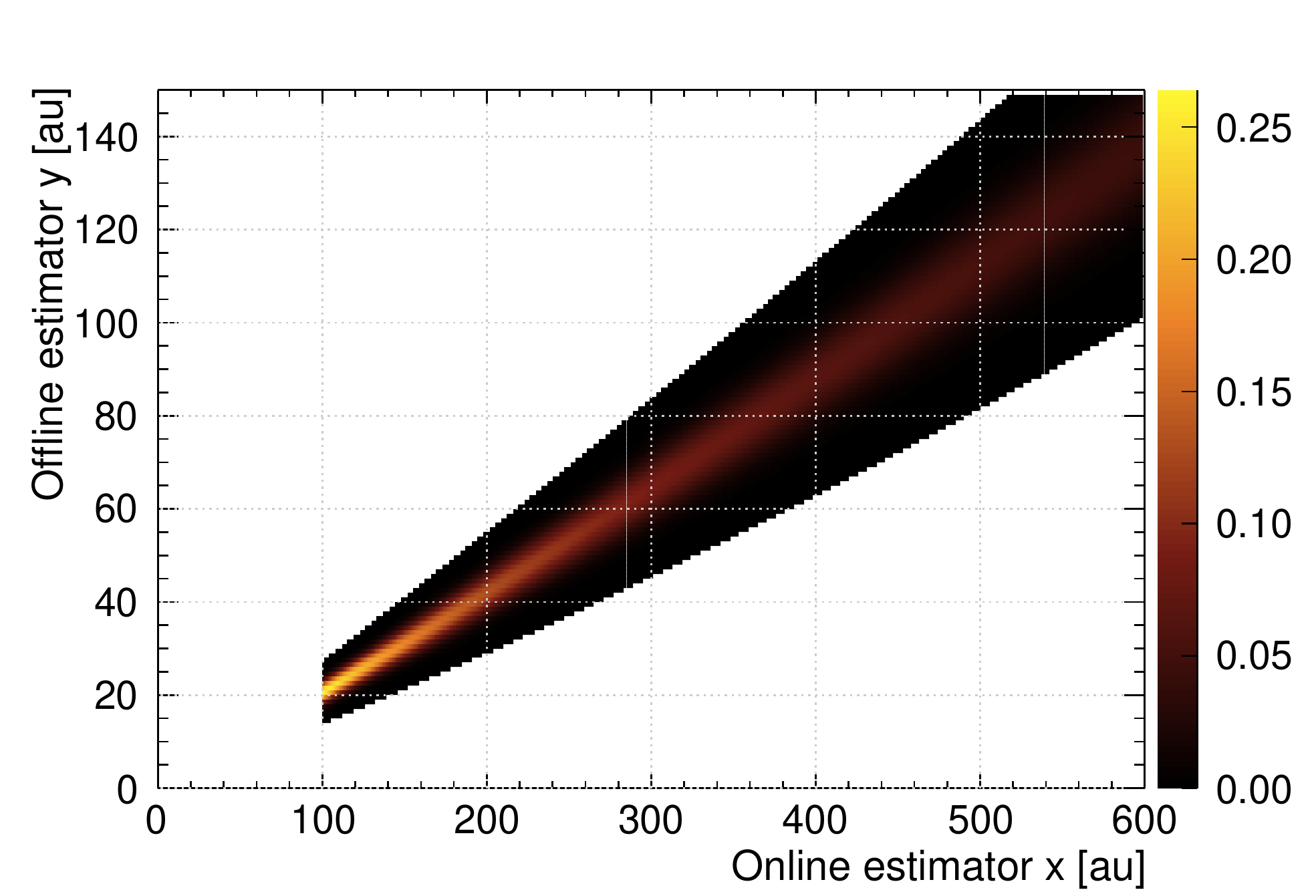}}\\
\subfloat[]{\includegraphics[width=0.75\columnwidth]{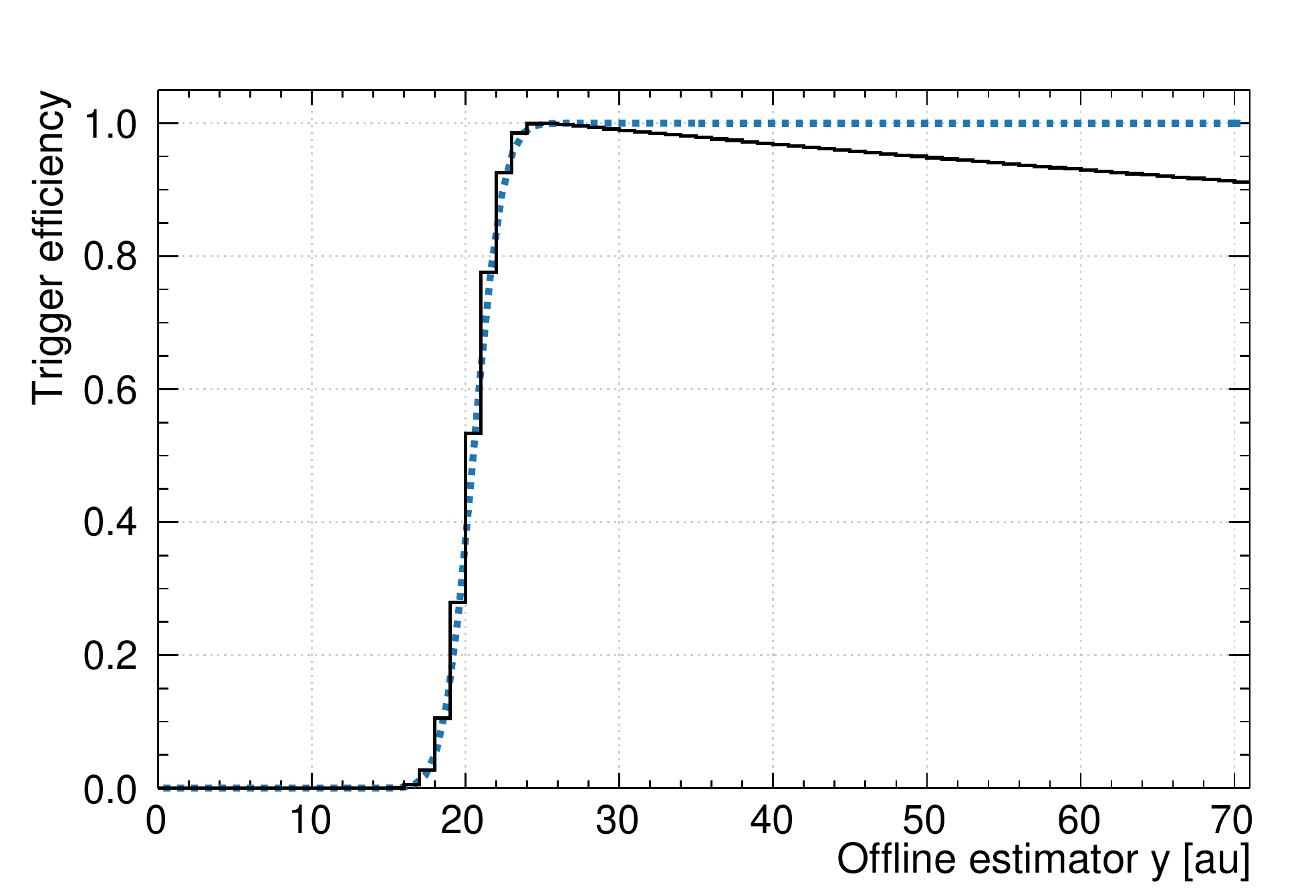}}
\caption{The Gaussian example with $\mu = 0.2y + 5\cdot10^{-5}y^2$ and $\sigma = 0.015y$. No plateau arises, hence the method is not applicable and panel b) does not represent the trigger efficiency.}
\label{fig:gausmodelm3s2}
\end{center}
\end{figure}

This method does not produce proper efficiency curves in all situations. Figs.~\ref{fig:gausmodelm2s3} and \ref{fig:gausmodelm3s2} show situations when it does not work, namely when at least one of the mean or the sigma functions are polynomials of level bigger than 1. In both figures, the trigger efficiency model from  Eq.~\eqref{eq:gausefficiency} is drawn to illustrate the differences.

\subsection{Skewed Gaussian example} \label{sec:skewgauss}
To show that this method does not work only for Gaussian distributions, we repeat the calculation for skewed Gaussian (also called exponentially modified Gaussian (EMG)) resolution functions:
\begin{linenomath*}
\begin{align}
I_x(x, y) = &\frac{\lambda\small(x\small)}{2} e^{\frac{\lambda\small(x\small)}{2}(2\mu(x)+\lambda(x)\sigma(x)^2-2y)} \nonumber \\ 
 &\cdot\erfc\big( \frac{\mu(x) + \lambda(x)\sigma(x)^2 - y}{\sqrt{2}\sigma(x)} \big) H(x-\theta_x)\label{eq:skewedg}
\end{align}
\end{linenomath*}

\begin{figure}[htb]
\begin{center}
\subfloat[]{\includegraphics[width=0.75\columnwidth]{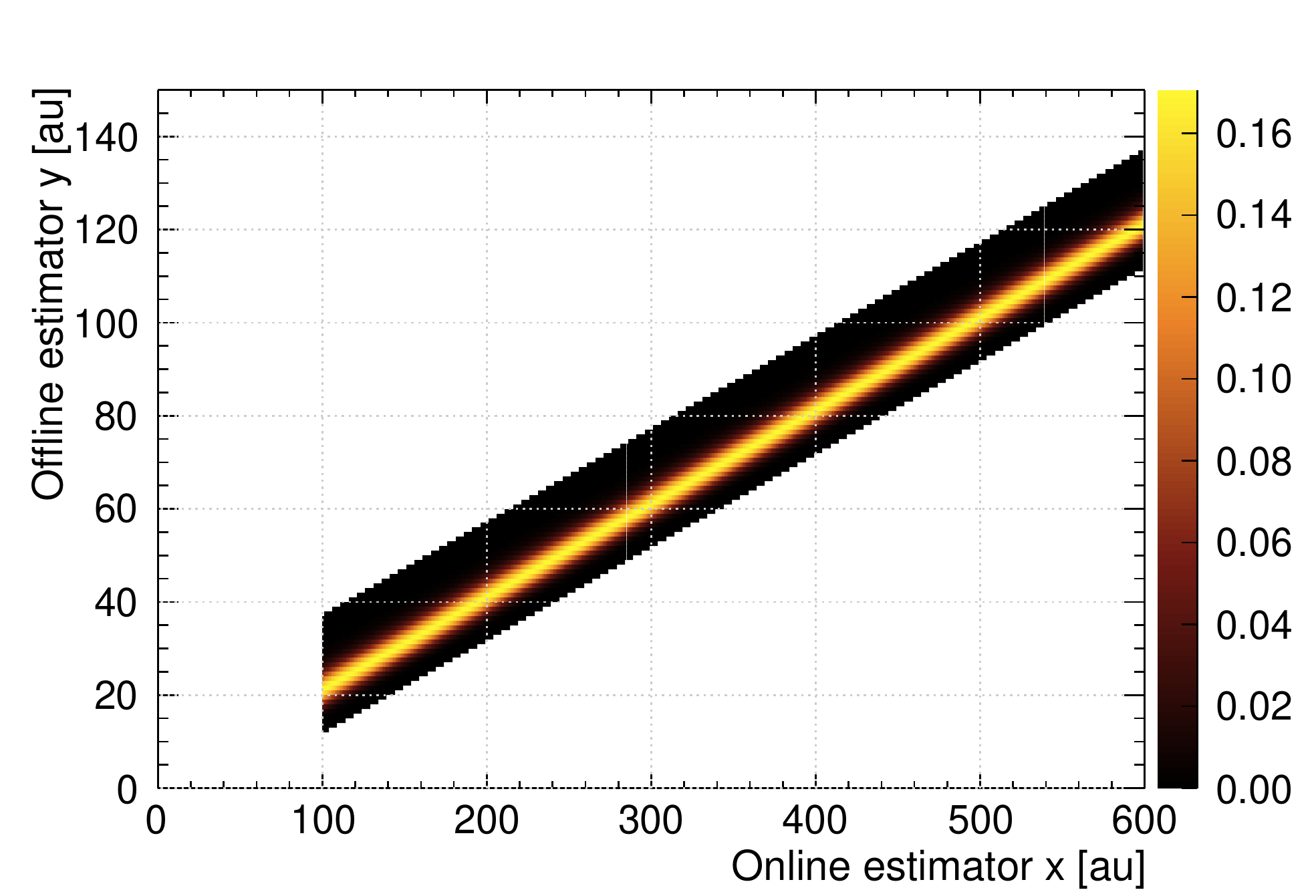}} \\
\subfloat[]{\includegraphics[width=0.75\columnwidth]{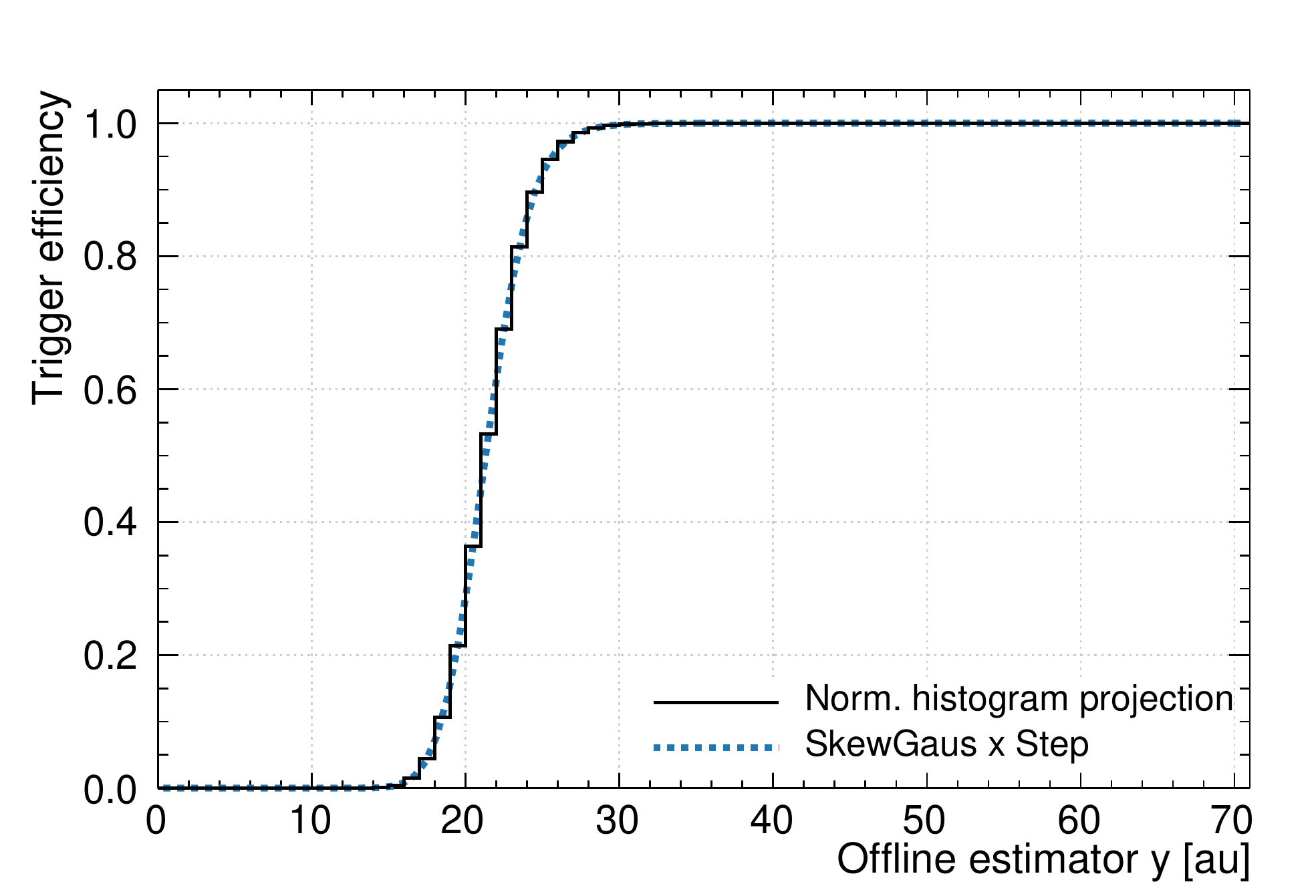}} \\
\caption{Skewed Gaussian example with $\mu = 0.2y$, $\sigma = 3$, and $\lambda = 0.7$. a) x vs y. b) The normalized y profile. Also shown (blue dashed) is a skewed Gaussian convoluted with a step function, and function parameters are taken at the trigger threshold.}
\label{fig:skewgausm2s1}
\end{center}
\end{figure}

In the simplest case, $\sigma(x) = \sigma$, $\lambda(x) = \lambda$, and $\mu(x) = a\cdot x$, shown in Fig.~\ref{fig:skewgausm2s1}. The y-axis projection without the trigger condition is then\footnote{The lower integration bound is set at 0 because of the definition of the EMG.} (see Appendix~\ref{app:EMG}).
\begin{linenomath*}
\begin{align}
\hat{I}_x(y; a, \sigma, \lambda) = &\int_{0}^\infty \frac{\lambda}{2} e^{\frac{\lambda}{2}(2ax+\lambda\sigma^2-2y)}\nonumber \\
 & \,\,\cdot\erfc\big( \frac{ax + \lambda\sigma^2 - y}{\sqrt{2}\sigma} \big) dx \label{eq:skewedgint} \\
	= &\frac{1}{a}
\end{align}
\end{linenomath*}

Again, a constant plateau height is expected in the absence of the trigger.

The analytic shape of the trigger turn-on curve is again obtained by including the trigger condition:
\begin{linenomath*}
\begin{align}
I_x(y; a, \sigma, \lambda, \theta_x) = &\int_{0}^{\infty}  I_x(x, y) H(x-\theta_x) dx \label{eq:skewgausintegral} \\
		= &\frac{1}{2a} \big[1 - e^{\frac{\lambda}{2}(2a\theta_x+\lambda\sigma^2-2y)} \nonumber \\
		 & \,\, \cdot \erfc(\frac{\sigma}{\sqrt{2}}(\lambda + \frac{a\theta_x-y}{\sigma^2}))  \nonumber \\
		 &\,\, + \erf(\frac{1}{\sqrt{2}\sigma}(y-a\theta_x))  \big] \label{eq:EMGCDF}
\end{align}
\end{linenomath*}
and dividing by the plateau height
\begin{align}
 \epsilon(y) = a \cdot I_x(y; a, \sigma, \lambda, \theta_x) \label{eq:EMGEfficiency}
\end{align}

Fig.~\ref{fig:skewgausm2s1} shows the EMG model for values of $\mu = 2y$, $\sigma = 3$, and $\theta_x = 100$. The turn on curve, Eq.~\eqref{eq:EMGEfficiency}, is drawn as well. Function parameters are taken at the trigger threshold (i.e. this is not a fit).

\begin{figure}[htb]
\begin{center}
\subfloat[]{\includegraphics[width=0.75\columnwidth]{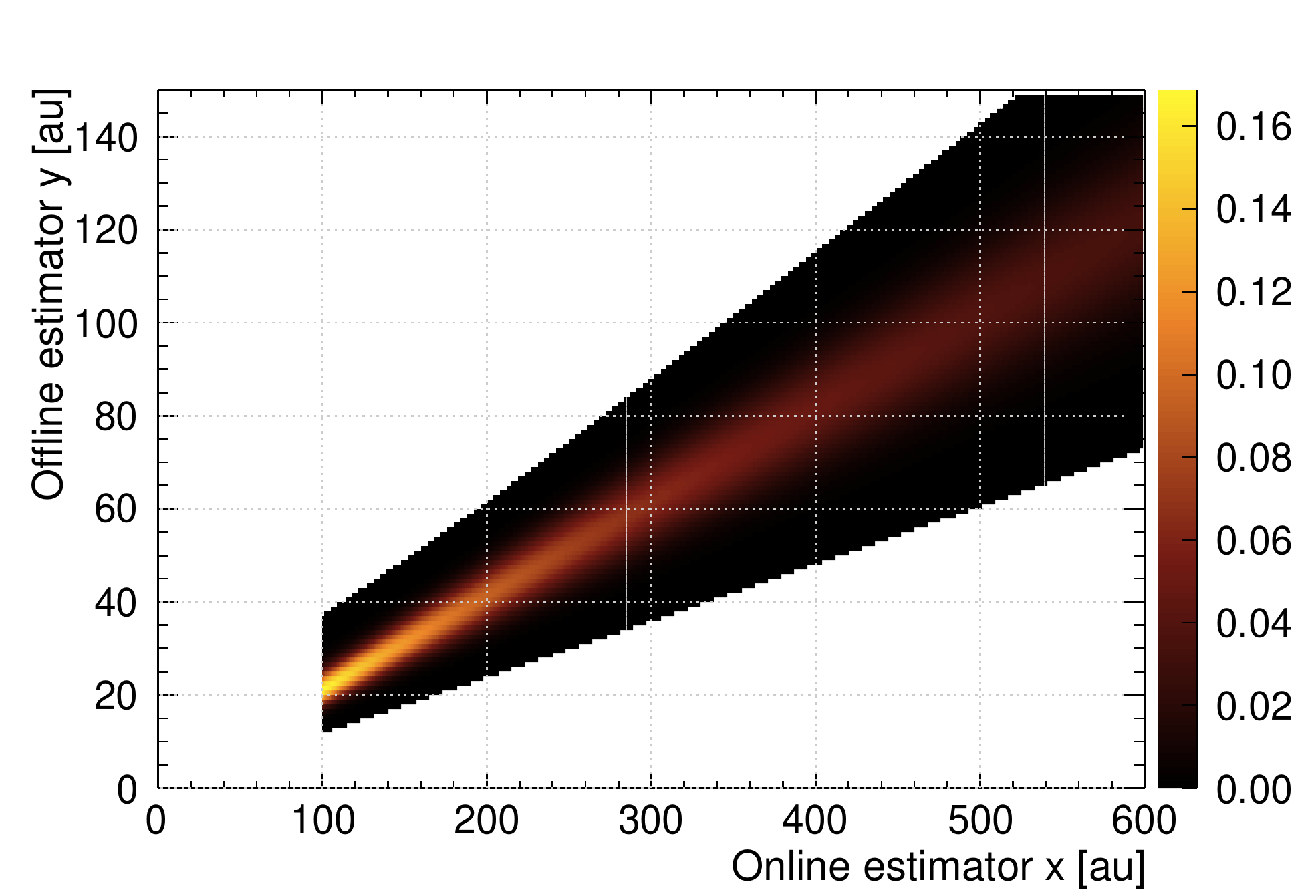}} \\
\subfloat[]{\includegraphics[width=0.75\columnwidth]{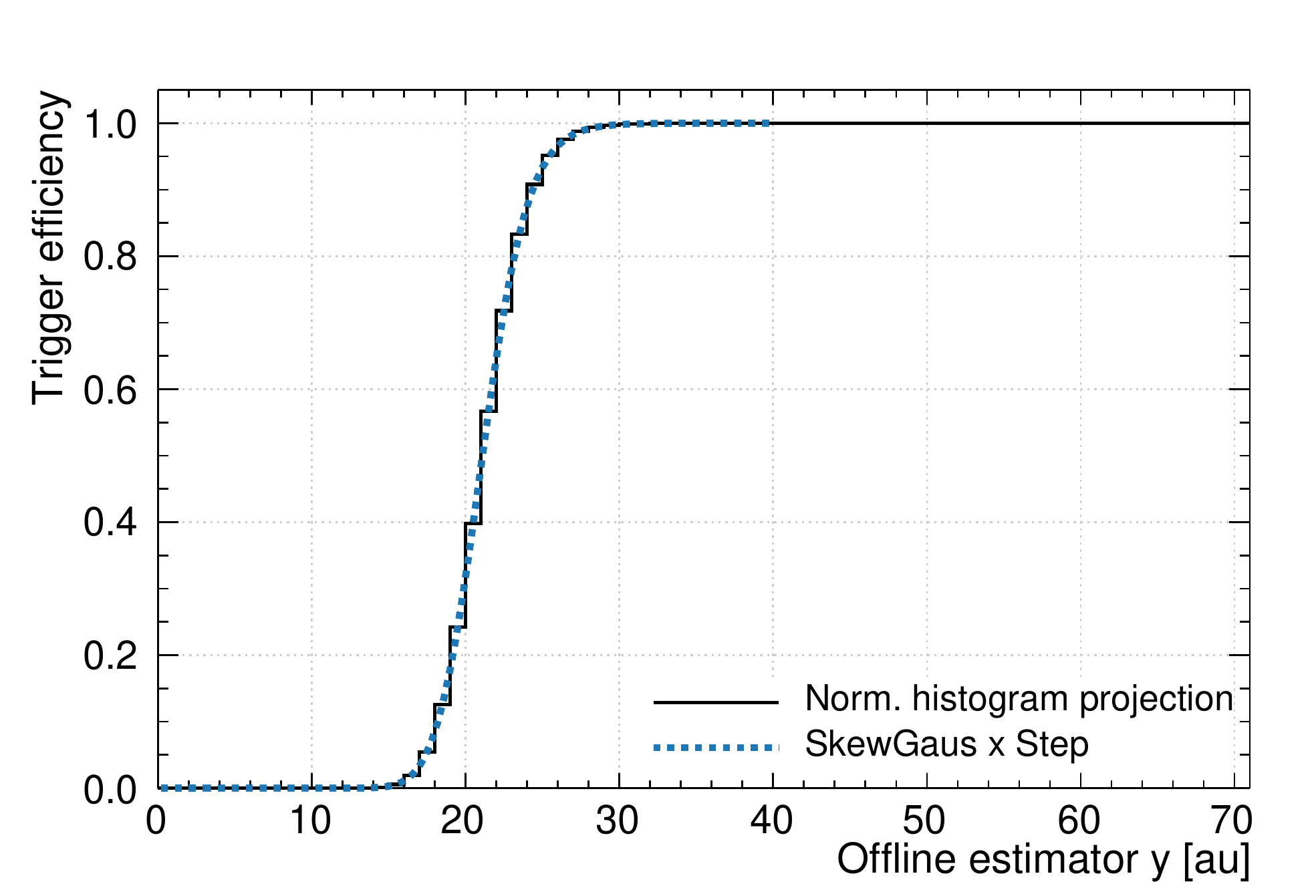}}
\caption{Skewed Gaussian example with $\mu = 0.2y$, $\sigma = 0.015y$, and $\lambda = 0.7$. a) x vs y. b) The normalized y profile. Also shown (blue dashed) is a fit to the curve with Eq.~\eqref{eq:EMGEfficiency}. }
\label{fig:skewgausm2s2}
\end{center}
\end{figure}

Fig.~\ref{fig:skewgausm2s2} shows the skewed Gaussian example for $\mu = 0.2y$, $\sigma = 0.015y$, and $\lambda = 0.7$. The efficiency curve again shows a plateau and the shape is described by Eq.~\eqref{eq:EMGEfficiency}, but here the parameters of the turn-on curve were fit out to $\mu = 19.8$, $\sigma = 2.0$, and $\lambda = 0.705$, so the parameters that determine the shape differ slightly from the parameters of the skewed Gaussian at the trigger threshold.

As previously seen in the Gaussian example, the method fails if $\mu$ or $\sigma$ are polynomials of order bigger than 1 in $y$. It also fails if $\lambda$ is not constant, though if the dependence of $\lambda$ on x is not strong, an approximately flat plateau region is obtained.

\section{Uncertainties}
The normalization of the spectrum in $x$ adds correlated uncertainties to the statistical uncertainties of each bin in the $I_x(y)$ histogram. Then, the plateau level must be fit out, introducing an uncertainty in the `true' number of events. The final efficiency curve or histogram comes with a complicated mixture of correlated and uncorrelated, statistical and systematic uncertainties. Furthermore, because the `true' number of events is only an estimate, the efficiency histogram can have values bigger than 1.

Some assumptions can be made to simplify the uncertainties. The uncertainty on the total number of events in each $x$ bin will always be smaller than the statistical uncertainty on the events in any (x, y) bin. Thus, the correlated uncertainties can be neglected in the $I_x(y)$ histogram. 

The efficiency histogram can be fit with an analytic function where the maximum value is constrained to $1$. Then, confidence regions can be obtained in the usual way by varying the fit parameters within their uncertainties.

The uncertainty on the plateau height must be minded as a systematic uncertainty.

\section{Discussion}
The method introduced here allows an estimation of the trigger efficiency turn-on curve without the use of calibration data. This method will have a larger uncertainty than a typical efficiency calibration for the same amount of data used. However, since `physics' datasets are often much larger than calibration datasets, this method can result in a more precise estimate.

The method works if a number of assumptions are true: i) The trigger bases the trigger decision solely on an online parameter $x$, and the value of $x$ is known for each event. ii) The trigger curve in $x$ is a step function; that is the efficiency is known to be 0 for $x < \theta_x$ and 1 for $x \geqslant \theta_x$.  iii) The existing data covers the full available parameter space in the trigger turn-on region, and far enough into the plateau region to estimate the plateau height. iv) The distribution of the offline parameter $y$ for events with the same $x$, $I(y;x=const)$, has the same functional form at all values of $x$ (or at least for all values of $x$ in the critical turn-on region and far enough into the plateau region that the plateau height can be obtained). v) The $I(x, y)$ histogram shows a linear dependence of $y$ on $x$, and the width of the $I(y;x=const)$ distribution is a polynomial of order $\leqslant 1$ in $x$. Non-Gaussian distributions will have additional requirements on the distribution shape parameters. These do not have to be explicitly determined - if this method produces a flat plateau region, the conditions are met.

Condition i) is typically met in particle physics experiments. We note that if the trigger estimator $x$ is not written to file for each event, it can often by re-constructed by programming an offline analysis algorithm that reproduces the trigger module algorithm.

Condition ii) must be met such that the result of this method is in fact a trigger efficiency. This method determines the efficiency curve of $y$ with regard to $x$, not the efficiency of $y$ with regard to the actual trigger. But if the trigger efficiency in $x$ is a step function with values of either $0$ or $1$, the curve obtained is the trigger efficiency in $y$. If the trigger efficiency in $x$ is not a step function, then it must additionally be obtained another way before the trigger efficiency in $y$ can be determined. A typical situation where this is the case would be one where the trigger is pre-scaled or otherwise known to approach a value different from one.

Condition iii) means that the physics data recorded must contain a sufficient number of events with values of $x$ near the trigger threshold. If it does not, presumably, the trigger turn-on curve is not of interest to begin with.

Condition iv) is the only one that is not trivial to verify based just on the physics data. An unchanging functional form of the $I(y;x=const)$ distribution is a reasonable assumption in most cases, but should if possible be checked using a traditional efficiency calibration approach. If $I(y;x=const)$ is known analytically, such that the shape of the turn on curve can be obtained by convolution with a step function, then the shape of the data should be well described by this analytic turn-on curve. If the shapes do not match, it would be an indication that condition iv) is not met.

We showed analytically that this method works for certain forms of Gaussian and skewed Gaussian resolution functions. We expect that this method will work for many realistic distributions $I(x,y)$, as long as $I(y;x=const)$ tends to 0 at both tails. This can be intuitively understood. At the integration borders of $y = -\infty$ and $y = \infty$, the distribution is 0. $x$ determines where inside the integration region the distribution peaks (through $\mu = \mu(x)$), but since the integration goes from minus to plus infinity, the location of the distribution on the $y$-axis is not relevant.

\section{Summary}

We have presented a method to obtain the trigger efficiency turn-on curve for a physics dataset. This method uses only the physics data itself, that is it does not require calibration data. It is based on several assumptions that are fulfilled for many types of experiments but at least one of which is difficult to verify without a calibration. Therefore, this method is particularly well suited to tracking the efficiency turn-on curve over time. It can be verified against calibration at any point of data recording and, once verified, be used to obtain the trigger efficiency curve over time, for example if calibration parameters drift faster than it is reasonable to record calibrations.

It can also be used to find out where full efficiency is reached, even if the precise shape of the turn-on is not reliable because the method was not verified. This can be useful when a dataset must be analyzed for which no other calibration is available, to at least find out in which region the recorded data is reliable.

\appendix 

\section{The skewed Gaussian} \label{app:EMG}
The generic skewed Gaussian probability density function (PDF) is
\begin{linenomath*}
\begin{align}
\text{emg}(z; \sigma, \lambda) = \frac{\lambda}{2} e^{\frac{\lambda}{2}(2\mu+\lambda\sigma^2-2z)}\cdot\erfc\big( \frac{\mu + \lambda\sigma^2 - z}{\sqrt{2}\sigma} \big) \label{eq:skewedgpdf}
\end{align}
\end{linenomath*}

The cummulative distribution function (CDF) is \cite{Haney:2011wf}
\begin{linenomath*}
\begin{align}
\text{EMG}(z; \sigma, \lambda) = &\int_{0}^z  \text{emg}(z; \sigma, \lambda) \label{eq:emgcdfintegral} \\
        = &\int_{0}^{\infty} \int_{-\infty}^{z-\tau} (\lambda e^{-\lambda\tau}) ( \frac{1}{\sqrt{2\pi}\sigma} e^{-\frac{(\beta-\mu)^2}{2\sigma^2}}) d\beta d\tau  \label{eq:emgcdfconv} \\
		= &\frac{1}{2} \big[1 - e^{\frac{\lambda}{2}(2\mu+\lambda\sigma^2-2z)}\erfc(\frac{\sigma}{\sqrt{2}}(\lambda + \frac{\mu-z}{\sigma^2})) \nonumber \\
		& + \erf(\frac{1}{\sqrt{2}\sigma}(z-\mu))  \big] \label{eq:EMGCDFz}
\end{align}
\end{linenomath*}
which in the limit for $z \rightarrow \infty$ is $1$.

We are dealing with Eq.~\eqref{eq:skewedgint}, which is reproduced slightly re-written here
\begin{linenomath*}
\begin{align}
\hat{I}_x(y;\sigma, \lambda) &= \lim_{x\rightarrow\infty} \int_{0}^x \frac{\lambda}{2} e^{\frac{\lambda}{2}(2ax+\lambda\sigma^2-2y)}\cdot\erfc\big( \frac{ax + \lambda\sigma^2 - y}{\sqrt{2}\sigma} \big) dx
\end{align}
\end{linenomath*}

Do a variable transformation $\hat{x} = ax$ then

\begin{linenomath*}
\begin{align}
\hat{I}_x(y; \sigma, \lambda) &= \lim_{x\rightarrow\infty} \int_{0}^x\frac{\lambda}{2} e^{\frac{\lambda}{2}(2\hat{x}+\lambda\sigma^2-2y)}\cdot\erfc\big( \frac{\hat{x} + \lambda\sigma^2 - y}{\sqrt{2}\sigma} \big) \frac{1}{a} d\hat{x} \label{eq:emgixtrans}
\end{align}
\end{linenomath*}

Eq.~\eqref{eq:emgixtrans} is in its form similar to Eq.~\eqref{eq:emgcdfintegral} if $z = -\hat{x}$ and $\mu=-y$. We see from Eq.\eqref{eq:emgcdfconv} that the integral is unchanged when going from $z \rightarrow -z$ and $\mu \rightarrow -\mu$ at the same time, hence\footnote{Note the change in the limit because the argument changed sign.} 
\begin{linenomath*}
\begin{align}
\hat{I}_x(y; \sigma, \lambda) = &\lim_{x\rightarrow-\infty} \frac{1}{2a} \big[1 - e^{\frac{\lambda}{2}(-2y+\lambda\sigma^2+2ax)} \nonumber \\
&  \cdot \erfc(\frac{\sigma}{\sqrt{2}}(\lambda + \frac{ax-y}{\sigma^2})) + \erf(\frac{1}{\sqrt{2}\sigma}(y-ax))  \big] \label{eq:EMGCDF2} \\
     = &\frac{1}{a}
\end{align}
\end{linenomath*}

The profile in y after applying the trigger threshold is given by Eq.~\eqref{eq:EMGCDF2} when setting $x = \theta_x$ instead of taking the limit.

\bibliography{TEPaperBib}
\bibliographystyle{unsrt}

\end{document}